\documentclass[draft, 12 pt, nofootinbib]{revtex4-1}
\usepackage{amsmath,amsfonts,amssymb}
\usepackage{graphicx}

\newcommand\vev[1]{\langle #1 \rangle}
\newcommand\ket[1]{| #1\rangle}
\newcommand\bra[1]{\langle #1|}
\newcommand\CO{{\cal O}}
\newcommand{\CT}{{\cal T}}

\begin{document}
\title{Quenches on thermofield double states and time reversal symmetry}

\author{David Berenstein}
\affiliation {Department of Physics, University of California at Santa Barbara, CA 93106}

\begin{abstract} 
In this paper we study a quench protocol on thermofield double states in the presence of time-reversal symmetry that is inspired by the work of Gao, Jafferis and Wall.
 The deformation is a product of hermitian operators 
on the left and right systems that are  identical to each other and that lasts for a small amount of time. 
 We study the linear dependence on the quench to the properties of the deformation under time reversal.
If the quench is time symmetric, then the linear response after the quench of all T-even operators  vanishes.
This includes the response of the energy on the left system and all the thermodynamic expectation values (the time averaged expectation values of the operators).
Also, we show under an assumption of non-degeneracy of the Hamiltonian that the entanglement entropy between left and right is not affected to this order. 
We also study a variation of the quench where an instantaneous deformation is given by an operator of fixed T-parity and it's time derivative. It is shown that the sign of the response of the Hamiltonian is 
correlated with the T-parity of the operator. We can then choose the sign of the amplitude of the quench to result in a reduction in the energy. This implies a reduction of the entanglement
entropy between both sides.
\end{abstract}
\maketitle

\section{Introduction}

The eternal black hole has been argued by Maldacena to be given by a thermofield double state \cite{Maldacena:2001kr} on two copies of a conformal field theory. In a standard description,
the two sides of the thermofield double do not interact with each other, and their only relation to each other is the entanglement pattern between them. Van Raamsdonk argued that 
entanglement between degrees of freedom of two theories can lead to topology change in black hole spacetimes \cite{VanRaamsdonk:2010pw}. Because the theories do not
interact with each other, it is impossible to pass a message from one side of the black hole to the other one.

Gao, Jafferis and Wall argued that there is a double trace deformation that can lead to the black hole wormhole geometry becoming traversable \cite{Gao:2016bin}. 
The idea is to modify the Hamiltonian by a so called double trace
deformation $\delta H\simeq \epsilon(t) \CO_L(t) \CO_R(t)$ for a brief period of time near $t=0$. We will call this the GJW protocol. This makes it possible for both sides to interact and in this way to be able to communicate with each other.
Maldacena, Stanford and Yang where further able to interpret this protocol as a version of quantum teleportation \cite{Maldacena:2017axo}. 

A natural question to ask is if the GJW protocol decreases the expectation value of the energy or the entanglement entropy of the state after the quench, thereby shrinking the area of the corresponding black hole. 
In this paper we will study this problem in general quantum systems under the assumption of having a time reversal symmetry $T$, where the Hamiltonian is time symmetric.  We consider first the case where $\epsilon$ is also time symmetric. 
We study quenches where $\CO$ is either $T$-even or $T$-odd. We find that under mild assumptions of non-degeneracy of the spectrum and the thermofield double state, that there is no change of the expectation value of either the energy or the entanglement entropy, 
so long as $\epsilon$ is time symmetric. In fact, we prove that this is true for any $T$-even Hermitian operator, as is the case of time averaged operators that would correspond to thermodynamic properties of the system. 

This shows that any such effect to linearized order will require us to break T-invariance, either in the pulse $\epsilon$, or in the choice of operators in the protocol. 

We also study a more general instantaneous quench where
\begin{equation}
\delta H \simeq \epsilon \delta(t) \CO_L(t) \dot \CO_R(t)
\end{equation}
Here, we find that there is a non-zero response for the left Hamiltonian to leading order. This follows from positivity properties of the amplitudes for states in the thermofield double state. 
 The sign of the response depends on the T-parity of $\CO$: if we change the T-parity of $\CO$ we change the sign of the response. Moreover, if 
an initial generalized thermofield double state is time symmetric and diagonal, we show that the response on the left and the right Hamiltonian  are the same.
Finally, if we use this effect to decrease the energy on the left state starting with the canonical thermofield double state, we must also reduce the entanglement entropy between both halves. This is because 
the entanglement entropy of the thermofield double state is maximal given the average energy of the left half of the system. 
 
The proofs are elementary and only require algebraic reality conditions of the matrix elements of operators,  as well as positivity conditions of amplitudes of states in the thermofield double state.

\section{Time reversal symmetry}

A system with time reversal operation has an anti-unitary operation $T$ such that $T^2=1$. For our arguments, we are assuming a system where any state has an even number of fermions excited to begin with, this way we can assume that there are no degeneracies for the 
eigenstates of the Hamiltonian \footnote{A more general setup would have states with odd fermion parity where $T^2=(-1)^F$. For these states, there are degeneracies and  issues with phases that would 
need to be addressed before the arguments in this paper could be applied.
}. 

Notice that if we have an eigenstate of $T$ with eigenvalue $1$,  $T \ket \psi= \ket \psi$, then it is true that $T(i\ket \psi)= -i T\ket \psi=-i \ket \psi$
changes the sign of the eigenvalue.
We will call the eigenstates of $T$ with eigenvalue $1$ real, and the ones with eigenvalue $-1$ imaginary. 
Any state can be decomposed into a real and imaginary part as follows
$$
\ket \psi = \left(\frac {1+T}2\right) \ket \psi +  \left(\frac {1-T}2\right)\ket \psi
$$
The Hilbert space can be spanned by complex combinations of  real states.
We say that the Hamiltonian is T-preserving if $T H T = H$. In this case, time reversal is a symmetry. If we evaluate the matrix elements of $H$ between real states, we get that 
\begin{equation}
\bra m T H T \ket n =  \bra m H \ket n = H_{mn} = \bra m \sum_ s H^*_{s n} \ket s = H^*_{mn}
\end{equation}
We find that the matrix elements of $H$ between real states are real.  In particular, the eigenvectors of $H$ can be taken to be real.

The same is true for any $T$-preserving operator $T{\CO} T= \CO$. Such an operator has purely real matrix elements between real states.
If $\CO$ is hermitian, it is realized by a real symmetric matrix in any real basis.
Similarly, for $T$-odd operators we get that $T{\CO} T=-\CO$, and the matrix elements of $\CO$ are purely imaginary in a real basis.
If furthermore $\CO$ is hermitian, then the matrix elements of $\CO$ are antisymmetric in any real basis.

\section{Double sided quenches in (generalized) thermofield-double states}

A thermofield double state is a state on a product Hilbert space of two copies of a physical system (left and right, or $L,R$ for short), which is of the form
\begin{equation}
\ket{TFD}= \sum_n \exp(-\beta E_n/2) \ket n_L \ket n_R
\end{equation}
where $\ket n$ are eigenstates of the energy operator $H_L$ acting on the left copy of the system.

We can use a shorthand notation where $\ket{TFD} \simeq \exp(-\beta H_L/2)$ as a matrix.
Here, for $T$-symmetry, we require that the basis $\ket n$ is real: this unambiguously determines the phase of the different amplitudes of the  thermofield double state. 
In \cite{Cottrell:2018ash} the authors use the antiunitary  $CPT$ operator to define the thermofield double state and study how to engineer the thermofield double state in simple systems.
For us, the presence of additional symmetries like $C,P$ lead to additional degeneracies that we want to avoid, so we will only study systems with $T$-symmetry
under the assumption that there are no degeneracies.

The total Hamiltonian for the system is $H_{tot} = H_L+H_R$, where $H_R$ has the same matrix elements of $H_L$, but on the right copy. 
If we evolve the system in time, we get that the amplitudes of the state in the energy basis develop phases
\begin{equation}
\ket{TFD (t) }=\sum_n \exp(-\beta E_n/2) \exp(-2 i E_n t) \ket n_L \ket n_R
\end{equation}
but the reduced density matrix $\rho_L= \sum \exp(-\beta E_n) \ket n \bra n$ stays unchanged.  The condition of $T$ invariance requires that the phases are aligned,
but moreover, we want the coefficients to be positive. This will be important later on.  
A thermofield double state is also annihilated by 
$H_L-H_R$.

We will say a generalized thermofield double state is a state that
is of a similar form, where it is a sum of products of (repeated) real states with {\em positive} real coefficients. That is, it is of the schematic form $\exp(-\beta H_{eff}/2)$
for some effective Hamiltonian that is time symmetric. This is the modular Hamiltonian for the density matrix $\rho_L$.  We can also require that it is  annihilated by $H_L-H_R$. For example a thermofield double for a microcanonical ensemble in some window would be of this form. We will not make a distinction between generalized and standard thermofield 
double states in what follows.

Notice the following fact. If $\ket{TFD}$ is a thermofield double state and $\phi$ is a T-even operator on a single copy of the system, then the matrix elements of $\phi$ are real. In particular, one checks that 
\begin{equation}
\vev{ \phi_L \phi_R}_{\ket{TFD}}  =  \sum_{mn} e^{-\beta (E_n/2+E_m/2)} \phi^2_{mn}  \geq 0\label{eq:sign}
\end{equation}
For the purposes of this paper we are interested on the sign of the response, so we do not have to normalize the states, otherwise, we should use an additional factor of the inverse of the partition function on the right.

This result is true regardless of if the thermofield double state is a canonical thermofield double or a general state. The sign depends on the positivity of the diagonal entries of the matrix $\exp(-\beta H_{eff}/2)$. 
The two T-even identical observables are positively correlated.
Similarly, if $\phi_L= \phi_R$ are T-odd, they are purely imaginary and we find that 
\begin{equation}
\vev{ \phi_L \phi_R}_{\ket{TFD}}  = \sum_{mn}  e^{-\beta (E_n/2+E_m/2)} \phi^2_{mn}  \leq 0
\end{equation}
so the correlation has the opposite sign. The T-odd observables are anticorrelated.

A double sided  quench \'a la Gao-Jafferis-Wall will be a unitary operator of the form
\begin{equation}
U_{Quench} = {\CT}\exp\left(-i  \int \epsilon(t) \phi_L(t) \phi_R(t)\right)
\end{equation}
acting on the thermofield double state. Here $\CT$ indicates time ordering as is appropriate for time dependent perturbation theory in the interaction picture.
We are interested first in time symmetric quenches where $\epsilon(t)=\epsilon(-t)$. The function $\epsilon$ is real and of very short duration. We also want it to be small (infinitesimal) and where $\phi_L(0)=\phi_R(0)$ are the same 
hermitian operators  acting on a single copy of the Hilbert space. The operator $\CO_L(t)$ is the time evolved operator of $\CO_L(0)$ under $H_L$ and similar for $\CO_R(t)$.   We will also require that they are both either $T$-even functions 
or both are $T$-odd \footnote{The GJW protocol uses a different convention of the relation between the left time and the right time which is  appropriate for global coordinates of the double sided black holes that results in an expression of the form $\CO_L(t)\CO_R(-t)$ rather than as above.}.

We will now study  the linearized response in $\epsilon$ of the (generalized) thermofield double state to the quench. 
For this, consider a $T$-even operator $W_L$ and consider the response to the quench from Kubo's formula.
\begin{equation}
\delta_\epsilon \vev{W_L} =-  \Im m ( \vev{[\int \epsilon(t) \phi_L(t) \phi_R(t), W_L]})_{\ket {TFD}}
\end{equation}
It is easy to argue that for a time symmetric $\epsilon$, the operator $\int \epsilon(t) \phi_L(t) \phi_R(t)$ is T-even.  Now, as argued in the previous section, the basis that diagonalizes the thermofield Hamiltonian $H_{eff}$ state is real and the thermofield double state itself is real.
 As such, the matrix elements of both $\int \epsilon(t) \phi_L(t) \phi_R(t)$ and $W$ are real in this basis. It immediately follows that $\delta_\epsilon W_L=0$, as the right hand side has no imaginary part. The condition of shortness of the pulse 
 is that $W$ should be measured after the quench, but at a small enough time that it can be replaced by $W(0)$. This depends on $W$ itself.
 
For example, this result pertains to the Hamiltonian $H_L$ and any time averaged (equilibrium) expectation value.  Time averaging removes the off-diagonal elements of an operator (so long as the Hamiltonian is non-degenerate) and gives rise to a T-even operator that is time independent.
Because of time translation invariance of these quantities, they can be measured after the quench and they will be $T$-even at any time, not only at $t=0$. Indeed, one can say based on this result that the double sided quench does not change the thermodynamic properties of the final state, to linearized order, with respect to 
the initial state after tracing over the right system degrees of freedom. 

Now, for any $T$-odd $W$ observable, we have that $\vev W_{TFD}=0$ by a very slight variant of the same argument as above: the expectation value is real as $W$ is hermitian, but the right hand side is purely imaginary.
In general, for these T-odd $W$, to linearized order the quench changes the expectation value. That is because the commutator in Kubo's formula is now purely imaginary and there is no constraint on its expectation value.
 
We will now argue that there is also no change in the entanglement entropy between both sides either,  to linearized order, for this type of quench.
To do this, we need to consider the thermofield double as a matrix $M$. The quench will change $M$ as follows
\begin{equation}
 M'  = M- i \int dt \epsilon (t) \phi_L(t) M \phi^T_R(t)+O(\epsilon^2)
\end{equation}
 where $\phi^T$ is the transpose matrix of $\phi$. The density matrix for the left system is given by $\rho_L= M' M^{\prime\dagger}$. To linearized order in $\epsilon$ we get that
 \begin{equation}
 \delta \rho_L = -i \int dt \epsilon (t) \phi_L(t) M \phi^T_R(t) M^\dagger +i M \int dt \epsilon (t) \phi(t)^*_R M^\dagger \phi^\dagger_L(t)
 \end{equation}
 The idea is to show that the right hand side is purely imaginary. This is straightforward to prove for an instantaneous quench: we use the reality properties of $M, \phi$ to do so. In particular in a  real basis we have that $M^\dagger =M$ is real, and $\phi$ is either purely real or purely imaginary.
 Thus the right hand side is purely imaginary.
 As such $\delta\rho$ is off-diagonal, after all, the initial $\rho$ is characterized by a hermitian operator, and it is real in a real basis.
  It follows that the eigenvalues of $\rho$ do not change to linearized order, by a standard perturbation theory calculation. This is where we require that the spectrum is non-degenerate. If the eigenvalues of $\rho$ only change to quadratic order in $\epsilon$, then the 
 entanglement entropy or any of the corresponding Renyi entropies are fixed at this order.  
 
We now need to show that this imaginary property is true in the general case. We concentrate our  attention on a fixed $t$ and its time reversed mirror time $-t$. Look at the first term on the right, summed over these two times
\begin{equation}
\phi_L(t) M \phi_R^T(t) M+ \phi_L(-t) M \phi_R^T(-t) M. \label{eq:t-symmetry}
\end{equation}
The result is clearly invariant under $t\to -t$. Now,  decompose $\phi_L(t) = \phi_e(t)+\phi_o(t)$ under even and odd parts with respect to $t$. Both of these are hermitian.
We do a similar decomposition for $\phi_R$.
It is clear that $\phi_E$ will have the same time reversal properties as $\phi(0)$, so it inherits its reality properties (either purely real or purely imaginary)  while the other piece will have the opposite
time reversal property. Because of the time symmetry of the full expression, only the even-even and odd-odd pieces contribute in the sum. They also have definite reality properties: either purely real or purely imaginary.  For each of these, we get a net reality property of the expression that is  the same as in the instantaneous quench problem.
We see that $\delta \rho$ is a sum (integral) of purely imaginary terms. As such, it is imaginary  and follows that it must be off-diagonal. By our argument for the instantaneous quench this implies that the entanglement entropy is not modified to linearized order.

\subsection{A double sided quench that changes the energy}

Now, we will concentrate on a slight variation of the double sided quench that leads to a change in the energy. The idea is that $\phi_L$ and $\phi_R$ should now have the opposite time reversal symmetry assignments, so that we can get a non-zero result. This is in lieu of considering cases where $\epsilon(t) = -\epsilon(-t)$, which would produce a time asymmetry in the quench protocol. We don't want the two sides to be unrelated. Instead, we will require that $\phi_R= \dot \phi_L$ and we will focus on an instantaneous quench for simplicity.  More general results follow if we use the same type of arguments that came after equation \eqref{eq:t-symmetry}. We will also focus on thermofield double states that 
are invariant under $H_L-H_R$. That is, they are diagonal in the basis that diagonalizes the Hamiltonian. Examples of such states are thermofield double states for microcanonical ensembles.
First, consider the response of the Hamiltonian on the left to the quench. By Kubo's formula we get that 
\begin{equation}
\delta_\epsilon \vev{H_L} = -\Im m ( \vev{[ \phi_L(0) \dot \phi_R(0), H_L]})_{\ket {TFD}}
\end{equation}
 Now, notice that $\dot\phi_L(0)= i [H_L,\phi_L(0)]$. We get this way that 
\begin{equation}
\delta_\epsilon \vev{H_L} =- \Im m ( \vev{[ \phi_L(0),H_L] \dot \phi_R(0)})_{\ket {TFD}}=- \Re e ( \vev{\dot \phi_L(0) \dot \phi_R(0)})_{\ket {TFD}}
\end{equation} 
If $\phi$ is T-even, then it is real, and $\dot \phi$ is purely imaginary. The right hand side would therefore be positive. This follows from the arguments that lead to equation \eqref{eq:sign}. If we choose the opposite assignment of T-parity,  we change the sign.  It is then clear that  we can choose
$\epsilon$ to lower the left energy depending on the T-parity of $\phi$.
Notice that so far we have not used the fact that the state is diagonal in the energy basis. This is used when we consider the response of the Hamiltonian on the right. We get that
\begin{equation}
\delta_\epsilon \vev{H_R} = -\Im m ( \vev{\phi_L(0) [\dot \phi_R(0), H_R])})_{\ket {TFD}}= -\Re e ( \vev{ \phi_L(0) \ddot \phi_R(0)})_{\ket {TFD}}
\end{equation} 
The matrix elements of $\ddot \phi_R(0)$ are given by $\ddot\phi_{nm}= - (E_n-E_m^2)\phi_{nm}$. 
Also, the matrix elements of the time derivative of the field are $\dot\phi_{nm} = i(E_n-E_m) \phi_{nm}$. We see that when we use the diagonal entries of the matrix $M\equiv \hbox{diag}(s_n)$, with $s_n\geq0$, we get that
\begin{equation}
\delta_\epsilon \vev{H_R}= \sum a_n a_m \phi_{nm}^2 (E_n-E_m)^2 =  \delta_\epsilon\vev{H_L}
\end{equation}
Notice that the result is symmetric with respect to both sides, only because  the state is diagonal in the energy basis. 

As an aside, if we do a  quench with the symmetric operator  $\phi_L \dot \phi_R+\dot \phi_L \phi_R= \partial_t(\phi_L\phi_R)$ we double the response, by the argument above. The instantaneous quench we described is computing half the result for $\epsilon(t)= \delta'(t)$, which is T-odd.
That is, our result implies that the quench with very short time $\epsilon(t)=-\epsilon(-t)$ has a definite sign for the response in the energy, and that this response changes sign if we change the sign of $\epsilon$. The sign depends on if the operators $\phi$ are T-even or T-odd. The result has opposite 
signs if we change the T-parity of $\phi$. 

As a corollary, if we use the canonical thermofield double, it is a state that maximizes the entanglement entropy given the expectation value of the energy. Since we can choose the sign of $\delta H$ to be negative by choosing the sign of $\epsilon$, we reduce the energy to linearized order. This implies that the entanglement entropy between both sides 
is also reduced to linearized order, by an amount that is at least as large as the difference in entropy between the two corresponding thermofield double states at their different fixed energies.  

That is, we get that for $\delta H_L<0$, the linearized response in the entropy is bounded below below by the thermodynamic relation $dE = T dS$, to be given by
\begin{equation}
|\delta S| \geq \beta|\delta H|
\end{equation}
where $\beta$ is the temperature. 
On the other hand, if we choose $\delta H$ positive, we get the opposite inequality, $\delta S \leq \beta \delta H_L$.
That is, we get that the change in entropy and the energy to first order are correlated exactly as in the thermodynamic relation. That is, we get 
that  for the canonical thermofield double states $\delta S = \beta \delta H_L$. 

For generalized thermofield double states we can not make this claim. Instead we also expect a response in the entanglement entropy.  
The fact that both are linear in the perturbation defines a generalized temperature by $\delta S= \beta_{eff} \delta H_L$, where $\beta_{eff}$ depends on
the choice of operator $\CO_L$ in the quench. We also have a thermodynamic relation that follows from the linear variation of the expectation value of the modular Hamiltonian $\delta S= \tilde \beta \delta H_{eff}$.

\section{Conclusion}

In this paper the response of various observables  to a ``double trace" quench applied to a generalized thermofield double state in the presence of time reversal symmetry was computed. 
 This is a quench where we deform the Hamitonian by a product operator 
$\delta H= \epsilon(t) \CO_L\times \CO_R$ where the operators are the same for a small amount of time encoded in $\epsilon$ being small and vanishing at both early and late times.
The quench protocol is inspired by the 
problem of traversable black holes from Gao, Jafferis and Wall. We saw that on time symmetric quenches with a double trace deformation, the response of the T-even observables vanishes if the operators involved in the quench have definite T-parity.
This includes all time averaged observables and even the entanglement entropy between both subsystems does not change to leading order.
This shows that to get a linear response to the quench for T-even operators one needs to break time symmetry either in the $\epsilon(t)\neq \epsilon(-t)$, or at the level of the choice of operators $\CO_L\neq \CO_R$.

We also saw that an instantaneous  quench where we deform by $\delta H=\epsilon \delta(t)\CO_L\times \dot\CO_R$ produces a linear response in the energy. The sign of the response depends on the T-parity of $\CO_L$ and changes sign if we change the T-parity of $\CO_L$.
This means we can change the sign of the response of the energy to lower the energy of the left. The result between left and right is symmetric if the generalized thermofield double state  is annihilated by the difference of the single sided Hamiltonian on both sides $H_L-H_R$.
It is also interesting to explore if this type of quench would also produce regenesis \cite{Gao:2018yzk}.

For some of the results, we depend on having a non-degenerate Hamiltonian and generalized thermofield double state. It would be interesting to generalize the calculations in this paper to more general cases involving degeneracies associated to additional symmetries of the system.
In particular, if we have a  charge that is odd under T-parity (as is common in angular momentum or electric charge), one can consider the problem of thermofield double states for systems with chemical potential. 
If the system is holographic, these setups would correspond to rotating black holes or charged black holes and one can expect that the response of the system can be classified further in terms of the charge assignments of the perturbations. The traversability of the corresponding black holes under double trace perturbations has been analyzed in examples in \cite{Caceres:2018ehr}.
In general, we did not use the `trace' counting that is appropriate to large N theories
in our results. It is not clear to the author how this property of the perturbation affects the discussion of the quench. 

In view of the vanishing results to linearized order in time symmetric quenches, it is also interesting to generalize the results in this paper to second order in the perturbation and check if it is possible to make statements of positivity of the response of various observables to second order.

\acknowledgements
The author would like to thank X. Dong, B. Freivogel, D. Marolf, M. Srednicki for many discussions. Work of D. B. supported by the Department of Energy grant DE-SC0019139.

\end{document}